\documentstyle[12pt]{article}
\begin{document}
\rightline{SU-ITP-903\ \ \ \ \  }
\rightline{October 28, 1991}
 \newcommand{\Psl}{\not\!\! P}
\newcommand{\dsl}{\not\! \partial}
\newcommand{\half}{\frac{1}{2}}
\def\a{\alpha}
\def\b{\beta}
\def\al{\aleph}
\def\g{\gamma}\def\G{\Gamma}
\def\d{\delta}\def\D{\Delta}
\def\e{\epsilon}
\def\et{\eta}
\def\z{\zeta}
\def\t{\theta}\def\T{\Theta}
\def\l{\lambda}\def\L{\Lambda}
\def\m{\mu}
\def\f{\phi}\def\F{\Phi}
\def\n{\nu}
\def\p{\psi}\def\P{\Psi}
\def\r{\rho}
\def\s{\sigma}\def\S{\Sigma}
\def\ta{\tau}
\def\x{\chi}
\def\o{\omega}\def\O{\Omega}
\def\lagr{{\cal L}}
\def\cd{{\cal D}}
\def\k{\kappa}
\def\tz{\tilde z}
\def\tF{\tilde F}
\def\ri {\rightarrow}
\def\cf{{\cal F}}
\def\pa {\partial}
%\begin{titlepage}
\vskip 2 cm
\begin{center}
{\large\bf GEOMETRY OF SCATTERING  \\
\vskip 0.5cm
AT PLANCKIAN ENERGIES }\\
\vskip 1.2 cm
{\bf R. Kallosh\footnote{On leave of
absence from: Lebedev Physical Institute, Moscow, 117924, USSR }}
\vskip 0.2cm
Physics Department, Stanford University\\
Stanford   CA  94305
\end{center}
\vskip 1.6  cm
\begin{center}
{\bf ABSTRACT}
\end{center}
\begin{quote}

\ \ \ \ \
We present an alternative derivation and geometrical formulation
of Verlinde topological field theory, which may describe
scattering at  center of mass energies comparable or larger than
the Planck energy.  A consistent truncation of 3+1 dimensional
Einstein action is performed using the standard geometrical
objects, like tetrads and spin connections. The resulting
topological invariant is given in terms of differential forms.
\end{quote}

%\end{titlepage}

\newpage

1. Recently a remarkable progress in understanding new aspects
of 3+1-dimensional gravity has been achieved by H. and E.
Verlinde \cite{VV}. Their work sheds a new light on the results
obtained by 't Hooft  on  scattering at center of mass energies
comparable or larger than the Planck energy.  In a series of his
papers  \cite{Hooft} it is shown that the high energy amplitudes
have a universal behavior,  similar to the behavior of two
dimensional string amplitudes. This result has not been well
understood and accepted by the scientific community.

Recent results of Verlinde stimulate new attempts to  understand
the whole complex of related ideas in a more deep way. They
have found that the high energy scattering is described by a
 topological field theory, which is a gauge-fixed
3+1-dimensional   Einstein theory,
 the quantization being performed under specific ``high energy
conditions". These conditions are attributed in  \cite{VV} to the
properties of the forward scattering at the center of mass energies
comparable or larger than the Planck energy. In  \cite{VV} the
classical Lagrangian has been described  in terms of  metric. For
performing the gauge fixing, however, it was necessary to
introduce some  vector fields in addition to  metric.  These vector
fields are analogous to the fluid velocity in fluid mechanics. The
final theory  of  high energy scattering  presented in \cite{VV} is
given in terms of a 3+1 dimensional topological theory. The
action, being a total derivative, is expressed through an integral
over the boundary. The resulting 3-dimensional action is very
simple, being quadratic in the dynamical variables. The physical
field configuration is subject
 to the constraint that the  vector fields are curl free in the 3+1
dimensional space (in the absence of matter).

The theory of high energy scattering presented in \cite{VV} is
extremely interesting. Unfortunately, however, the geometrical
meaning of the additional  vector fields  introduced  in \cite{VV}
is somewhat obscure. In addition to this problem, the BRST
quantization which has been performed in \cite{VV}, was not
really a standard one. After the gauge fixing, the action of ghost
fields together with some part of the classical action
(proportional to the constraint on vector fields) have been
excluded from the final theory.

The purpose of the present paper is to represent the effective
high energy theory in a  completely geometrical way, without any
problems with the gauge fixing.  We will show that  it is not
necessary to perform BRST quantization in order to obtain the
corresponding topological field theory under the high energy
conditions used in \cite{VV}:  It is quite sufficient to make a
consistent truncation of the classical theory.  We will start with
3+1-dimensional Einstein theory using   tetrads and spin
connections instead of metric.  This will allow us  to avoid
introducing any auxiliary variables like the  above mentioned
vector fields. We are going to take another advantage by working
in the so called 1.5 formalism \cite{FV,PN}, where the spin
connections $w$ are the functions of  the tetrads $e$. The
functions $w(e)$ are such that the equation of  motion for spin
connections is solved.

By the consistent truncation of a gauge theory we mean the
following.
 We start with a gauge invariant action
\begin{equation}
S(\phi, \F)\ ,
\label{full}\end{equation}
depending on the set of gauge fields $\phi^i, \F^a$ with the
gauge symmetry $\d \phi^i = R^i_\a(\f, \F) \xi^\a$
 and $\d \F^a = R^a_\a(\f, \F) \xi^\a$.  One can perform the
consistent truncation as follows. Suppose we want to exclude the
fields $\F^a$. To make this consistent one must require that the
classical field equation for these fields is satisfied. The new
action is \begin{equation} \tilde {S}(\phi) \equiv S(\phi, \F(\f))\
, \label{trun}\end{equation} where the function $\F^a(\f)$ is a
solution of the equation   \begin{equation} \frac{\partial S(\phi,
\F)}{\partial\F^a} = 0 \ . \end{equation}
The action  $\tilde {S}(\phi) $ defined by equation (\ref{trun})
depends on less variables than the original action (\ref{full}).
However, it is still gauge invariant under the gauge
transformations of the smaller set of fields, \begin{equation} \d
\tilde {S}(\phi) = \frac{\partial \tilde {S}(\phi)}{\partial\f^i}
R^i_\a(\f, \F(\f))\xi^\a = 0 \ .   \label{sym} \end{equation} To
prove that the truncated action is indeed symmetric, we can find
its variation using eq. (\ref{trun}).   \begin{equation} \d \tilde
{S}(\phi) = \d S(\phi, \F(\f)) =   \{ \frac{\partial S (\phi, \F)
}{\partial\f^i}   R^i_\a(\f, \F) \xi ^\a+  \frac{\partial S
(\phi, \F) }{\partial\F^a}
R^a_\a(\f, \F) \xi ^\a   \} _{\mid \F= \F(\f)} = 0 \ .
\label{gi}\end{equation}
The expression in curly braces is equal to zero due to the
 invariance of the full action even before we substitute $ \F(\f)$
for $\F$ . After this substitution the second term in curly braces
vanish,  and we are left with the first term only, which proves
the  gauge symmetry of the truncated action, expressed by eq.
(\ref{sym}). \vskip 0.6 cm

2. Consider the 3+1-dimensional Einstein-Hilbert action in the
first order formalism, \begin{equation}
S^1 = \int e^m \wedge e^n \wedge R^{pq}(w) \e_{mnpq}\ ,
\label{E-H}\end{equation}
where the tetrad forms $e^m = dx^\m e_\m^m$  are $SO(3.1)$
tangent space
 vectors, $m = 0,1,2,3 ,\quad \m = 0,1,2,3 $. The curvature forms
\begin{equation}
R^{pq} = (d w + w \wedge w)^{pq}
\end{equation}
are $SO(3.1)$ tangent space tensors. The spin connection forms
$w^{pq} = dx^\m w_\m^{pq}$ are the connections for the local
Lorentz group  acting in $SO(3.1)$ tangent space.  If the spin
connections are  independent variables in the Lagrangian in
addition to tetrads, this is the first order formalism. The
classical equation for spin connections is \begin{equation} D e^m
\wedge e^n \e_{mnpq} = 0\ ,  \label{con}\end{equation} where the
torsion-free condition \begin{equation} T^m = D e^m  = d e^m + w
^m_n \wedge e^n = 0  \label{tor}\end{equation} solves equation
(\ref{con}). The solution to equations (\ref{con}), (\ref{tor}) is
given by \begin{equation} w_{\m mn}(e) =  \half e_m^\n ( \pa_\m
e_{n\n} - \pa_\n e_{n\m} )- \half e_n^\n ( \pa_\m e_{m\n} -
\pa_\n e_{m\m} ) - \half e_m^\r  e_n^\s  ( \pa_\r e_{p\s} - \pa_\s
e_{p\r} )  e_\m^p  \ .  \label{sol}\end{equation}
One gets the second order formalism if one substitutes  $w_{\m
mn}(e) $ from eq. (\ref{sol}) for $w_{\m mn}$ in the
Einstein-Hilbert  action (\ref{E-H})  and expresses the result as
the function of  tetrads only, \begin{equation} S^2(e) = \int e^m
\wedge e^n \wedge R^{pq}(w(e)) \e_{mnpq}\ .
\label{2}\end{equation} This is the simplest example of the
consistent truncation of the classical action  described above.
The action in the second order formalism (\ref{2}) depends on the
smaller set of fields than the action in the first order formalism
(\ref{E-H}). Still the action (\ref{2}) has all gauge
 symmetries as the action (\ref{E-H}): 3+1 general covariance in
the curved space
 and $SO(3.1)$ local Lorentz symmetry in the tangent space. All
these symmetries are realized in terms of tetrads only. The proof
that the consistently truncated theory (\ref{2})  is generally and
Lorentz covariant is a particular example of the proof given in
eqs. (\ref{sym}), (\ref{gi}) for arbitrary theory.

The 1.5 formalism is a mixed case. It has been used for the first
time by Fradkin and Vasiliev \cite{FV} in their discovery of
$SO(2)$ supergravity. This formalism is described in detail in
the review \cite{PN}. The 1.5 formalism action depends on tetrads
and on spin connections. The spin connection, however, satisfies
 the equation of motion given in eq. (\ref{sol}).
\begin{equation}
S^{1.5}(e, w(e)) =  \int e^m \wedge e^n \wedge R^{pq}(w(e))
\e_{mnpq}\ . \label{1.5}\end{equation}
\vskip 0.3 cm

3. According to Verlinde, we choose our $x$-axis along the  beam
of high energy particles and introduce the notation $x^\a \equiv
(t,x)$ for the longitudinal coordinates and $y^i \equiv (y,z)$ for
the transversal coordinates. The momenta of particles in the
$x^\a$-plane are of Planckian magnitude and the transversal
momenta are negligible when we are dealing with forward
scattering. The 3+1-dimensional tangent space is also given by
$SO(1.1)$ vectors $dx^\m e_\m^a = e^a, \quad a=0,1$,  and by
$SO(2)$ vectors $dx^\m e_\m^I = e^I, \quad I=2,3$. The full
exterior derivative operator $d$ now consists of longitudinal and
transversal parts:  \begin{equation} d = dx^\m \pa _\m = d^{lg} +
d^ {tr}\ , \quad  d^{lg} = dx^\a \pa _\a\ , \quad  d^{tr} = dy^i \pa
_i \ .   \label{form}\end{equation}  Thus, all original indices are
split as follows: $\m = \a, i ,\quad m = a, I $.  With this geometric
setup it is natural to try to perform a consistent truncation of the
3+1 gravity to  the smaller system without non-diagonal terms in
tetrads, i.e. to exclude from the theory the variables $e_\a^I$ and
$e_i^a$ by solving  equations of motions for them,
  \begin{eqnarray} \e^{\a \n\r\s} \e_{I npq} e_\n^n R
_{\r\s}^{pq}(w(e)) &=& \e^{\a \b ij} \e_{I abJ} e_\b^a R _{ij}^{bJ}
+ 2 \e^{\a i\b j} \e_{I Kab} e_i^K R _{\b j}^{ab} = 0\ , \nonumber\\
 \e^{i \n\r\s} \e_{a npq} e_\n^n R
_{\r\s}^{pq}(w(e))  &=& \e^{i j\a \b} \e_{a IbJ} e_j^I R
_{\a\b}^{bJ} + 2 \e^{i\a j\b} \e_{a bIJ} e_\a^b R _{j\b}^{IJ} = 0\ .
\label{cons}\end{eqnarray}
 Note, that to get these
equations we have varied the action $S^{1.5}(e, w(e))$ (\ref{1.5})
taking into account its  explicit dependence on tetrads. The
variation over the dependence on the tetrads through spin
connections $w(e)$ drops from the equations  of motion since the
corresponding term is \begin{equation} \frac{\pa S(e, w(e))}{\pa
w_\m^{mn}} \frac{\pa w_\m^{mn}}{\pa e_\s^p} = 0\ .
  \end{equation}
This term vanishes since in the 1.5
formalism $w$ satisfies equation
  \begin{equation} \frac{\pa S(e, w(e))}{\pa
w_\m^{mn}} = 0\ .
 \end{equation}

We will consider the solutions of classical
 equations  (\ref{cons}) at $e_\a^I =0
$ and $e_i^a = 0$. The truncated action is given in terms of
truncated tetrads (diagonal zweibeins ) and truncated spin
connections. The truncated tetrads are:
 \begin{equation} e^m_{trun} = \{e^a = dx^\a e_\a ^a , \quad  e^I =
dy^i e_i ^I  \}\ .
 \label{tt}\end{equation}
The truncated spin connections are functions of  zweibeins
$e_\a ^a, e_i ^I $ , given by eq. (\ref{sol}) at $e_\a ^I =0 $ and
$e_i ^a = 0$. The resulting truncated action is
\begin{equation} S_{trun}(e, w(e)) =  2  \int \left( e^a \wedge e^b
\wedge R^{IJ}(w(e)) +
 e^I \wedge e^J\wedge R^{ab}(w(e))
- 2 e^a \wedge e^I \wedge R^{bJ}(w(e)) \right)\e_{ab} \e_{IJ}\ .
\label{tr}\end{equation}
 This truncation is consistent only when the truncated tetrads
and connections satisfy classical eqs. (\ref{cons}).  At this point
we will add to the system described above the conditions which H.
and E. Verlinde attribute to the properties of the forward
scattering at Planckian energies. Note, that  until this point our
treatment of the 3+1 gravity  which led us to eqs.  (\ref{tr}),
(\ref{cons}) was exact, no approximations have been done yet.
\vskip 0.6 cm
4. The high energy conditions derived in \cite{VV} in terms of
metric variables are given by the equations \begin{eqnarray}
\pa_\a h_{ij} &=&0\ ,\nonumber\\
R_g &=&0\ ,
\label{high}\end{eqnarray}
where $h_{ij}$ is the transverse metric $h_{ij} = e_i ^I e_{jI}$
and $R_g$ is  the scalar curvature of the longitudinal space build
from the metric  $g_{\a \b} = e_\a ^a  e_{\b a}$. The solutions to
these conditions are taken in the form \cite{VV} \begin{eqnarray}
 h_{ij} &=& h_{ij} (y)\ ,\nonumber\\
g_{\a \b } &=& \eta_{ab} \pa _\a X^a \pa_\b X^b \ .
\label{string}\end{eqnarray}
We can reformulate both equations (\ref{high}) in terms of our
truncated tetrads (\ref{tt})
as one condition
\begin{equation}
d^{lg} e^m_{trun} =  0\ .
\label {V}\end{equation}
Note that the detailed form of this condition is
\begin{equation}
d^{lg}e^a = dx^\a \pa _ \a dx^\b e_\b ^a = 0  , \quad
d^{lg} e^I = dx^\a \pa _ \a dy^i e_i ^I = 0\ .
\label{tthigh}\end{equation}
The solution to high energy constraint (\ref{V}) is
\begin{eqnarray}
e^a &=& d^{lg} X^a(x^\a, y^i )\ ,\nonumber\\
 e_i ^I &=&  e_i ^I (y^j)\
\label{planck}\end{eqnarray}
where $X^a$ is some $SO(1.1)$ tangent vector zero form.
\vskip 0.3 cm

5. Our next step is to constrain our consistent truncated action
(\ref{tr}),   with the variables satisfying  eqs. (\ref{cons}),  by
the  high energy constraint (\ref{V}). This is quite
straighforward.  We will look for the spin connections which
simultaneously solve  the following system of equations:

i) the consistency condition for the truncation of tetrads  to
diagonal zweibeins, given in equations (\ref{cons})

 ii) the torsion free condition (\ref{tor}), (\ref{sol}) for
tetrads  $e_\m^m$ truncated to diagonal zweibeins, satisfying
eq.  (\ref{tthigh}) .

The solution to this system of equations is
\begin{eqnarray}
w^{ab} &\equiv & dx^\a w_\a^{ab} + dy^i w_i ^{ab} =  0\ ,
\nonumber\\ w^{IJ} &\equiv & dx^\a w_\a^{IJ} + dy^i w_i ^{IJ} =
 dy^i w_i ^{IJ}\ ,\nonumber\\
w^{aI} &\equiv & dx^\a w_\a^{aI} + dy^i w_i ^{aI} =
dx^\a w_\a^{aI} = e^{Ii} \pa_i e^a = d^{lg} e^{Ii} \pa _i X^a\ ,
\label{w}\end{eqnarray}
where
\begin{equation}
d^{lg} w^{aI} = (d^{lg})^2 e^{Ii} \pa _i X^a = 0
\ .
\end{equation}
In particular, our solution includes the condition $w_i^{ab}= 0$.
This is the exact
 counterpart to Verlinde constraint $\pa^{[\a} V_i^{\b]} = 0$
translated from the fluid mechanics language to the geometric
language of spin connections. The only non-zero components of
the zweibein-compatible spin connections  consistent
 with the high energy constraint are
\begin{eqnarray}
w_{i IJ} &=& \half e _I^j ( \pa_i  e_{Jj} - \pa_j e_{Ji}) -
\half e _J^j ( \pa_i  e_{Ij} - \pa_j e_{Ii}) - \half e_I^k e_J^l
( \pa_ik e_{Kl} - \pa_l e_{Kk } )e_i^K\ ,
\nonumber\\
w_{\a aI} &=&e^{Ii} \pa _i e^a_\a = e^i_I \pa _i \pa_\a X^a   \ .
 \label{nw}\end{eqnarray}
These solutions for spin connections imply
 that some components of curvature tensors vanish,
 \begin{equation}
R^{aI}_{ij} = R^{ab}_{\a i} = R^{aI}_{\a\b} = R^{IJ}_{i\a} = 0\ ,
\end{equation}
which solves eqs. (\ref{cons}).

Note, that the conditions
 \begin{equation}
w_\a^{ab} = w_\a^{aI} = w_\a^{IJ} = 0
\end{equation}
are the consequences of the high energy conditions and
torsion-free condition. The vanishing of $w_i^{ab}$,
 \begin{equation}
w_i^{ab}  = 0 \ ,
\end{equation}
comes from the solution of classical equations  (\ref{cons})
when the high energy conditions are are already imposed.

The curvature tensors which enter the truncated
action (\ref{tr}) are the following. In the first term we have the
curvature of the  transverse space  \begin{equation} R^{IJ}_{ij} =
R^{IJ}_{ij}(w_k^{KL})\ . \end{equation}
In the second term of (\ref{tr}) we have
\begin{equation}
R^{ab}_{\a\b} = w_{\a I}^a  w_{\b }^{Ib}  -   w_{\a I}^b  w_{\b }^{Ia}
\ . \label{R}\end{equation}
The third term contains
\begin{equation}
R^{aI}_{\a i} = D_i w_\a^{bI} \ .
\end{equation}

Now that the consistency condition for truncation has been
solved
 with the high energy constraints being taken into account, the
Einstein-Hilbert action has the same 3 terms as the truncated
action:  \begin{equation} S_{trun}^{Planck} = 2  \int ( e^a \wedge
e^b \wedge R^{IJ} +
 e^I \wedge e^J \wedge R^{ab}
- 2 e^a \wedge e^I \wedge R^{bJ})\e_{ab} \e_{IJ} \ .
\label{pt}\end{equation}
However, now we have in addition the consistency and the high
energy conditions: \begin{eqnarray}
d^{lg} e^m =
d^{lg} w^{mn}  = w^{ab} = d^{lg} R^{mn} = 0 \ ,\qquad m=0,1,2,3,
 \quad a=0,1.
\end{eqnarray}
The property of each term in this action to be a topological
invariant follows from the simple fact  that each term can be
represented as $\int d^{lg}$ of something. \begin{equation}
S_{trun}^{Planck} = 2  \int d^{lg} \{( X^a \wedge e^b \wedge R^{IJ}
+
  e^I \wedge e^J \wedge e^{Ki} \pa_i X^a w_K^b
-2  X^a \wedge e^I \wedge R^{bJ} ) \e_{ab} \e_{IJ} \}  = T\ .
\label{T}\end{equation}
In deriving equation (\ref{T}) we have used eqs. (\ref{planck}),
(\ref{w}) and (\ref{R}).
Being a total divergence, the action (\ref{pt}) is still a gauge
symmetric action: It  is general covariant and Lorentz covariant
for the gauge transformations which vanish at the boundary. This
property is in a complete agreement with our definition of a
consistent truncation of a gauge theory. The consistently
truncated gauge action must be gauge symmetric, and it is gauge
symmetric in our case!

\vskip 0.6 cm
6. If we denote the boundary values of $X^a$ by $\overline {X}^a$,
the 3+1 dimensional topological term (\ref{T}) can be presented
as an integral over the 3-dimensional boundary,
 \begin{equation}
T = \int d\tau \int dV^{tr} \e_{ab} \dot {\overline {X}}^a
(\triangle - R)_{tr}  \overline {X}^b \ ,
\label{action}\end{equation}
where $\tau$-time parametrizes the coordinates $x^\a(\tau)$ on
the boundary and $dV^{tr}$, $\triangle_{tr}$ and $R_{tr}$ are
the  volume of  integration, the scalar Laplacian and curvature in
the   transversal $y$-space. This is the action derived in
\cite{VV}. This action upon quantization leads to the
fundamental equal-$\tau$-time commutation relations
\begin{equation} [X^a (y_1), X^b (y_2)] = i \e_{ab} f(y_1, y_2)\ ,
\label{comm}\end{equation} where $f$ is the Green function
defined by the operator $(\triangle - R)_{tr} $. This commutator
was suggested before by 't Hooft \cite{Hooft} and has been related
to a new quantum gravitational uncertainty principle.

Our present formulation allows  to give a clear explanation of the
origin of this  commutator. We have started with the classical
Einstein-Hilbert action and performed a completely  consistent
truncation of a gauge system. We have added the high energy
constraint of Verlinde, which in our notations  takes a very
simple form (\ref{V}), $d^{lg} e^m_{trun} =  0$. The part of the
classical truncated Einstein-Hilbert action which does not vanish
when the constraint  (\ref{V}) is imposed, is given by the action
(\ref{action}), which leads to the commutation relations
(\ref{comm}). Thus, eq. (\ref{V}) is the only assumption behind
the uncertainty principle discussed above. \vskip 0.6 cm In
conclusion, we have derived the recent results of \cite{VV} in the
geometric language,  which is most appropriate for the
description of the problem of the forward scattering  at the
Planckian energies.

\vskip 0.6 cm
It is a pleasure to thank J. Russo, L. Susskind and L. Thorlacius
for valuable discussions. This work was supported in part by NSF
grant PHY-8612280 and by the John and Claire Radway
Fellowship in the School of Humanities and Sciences at  Stanford
University.

\end{document}